\documentclass{epl}
\newcommand{\be}{\begin{equation}}
\newcommand{\ee}{\end{equation}}
\newcommand{\bea}{\begin{eqnarray}}
\newcommand{\eea}{\end{eqnarray}}

\def\(({\left(}
\def\)){\right)}
\def\[[{\left[}
\def\]]{\right]}

\title{Disorder chaos in spin glasses} \author{Florent Krz{\c{a}}ka{\l}a\inst{1}
  \and Jean-Philippe Bouchaud\inst{2,3}} \institute{
  \inst{1} Laboratoire P.C.T., UMR CNRS 7083, ESPCI, 10 rue Vauquelin, 75005 Paris, France \\
  \inst{2} S.P.E.C., Orme des Merisiers,  CEA Saclay, 91191 Gif sur Yvette Cedex, France\\
  \inst{3} Science \& Finance, Capital Fund Management, 6-8 Bd Haussmann, 75009 Paris, France.\\
}


\begin{document}
\maketitle

\begin{abstract}
  We investigate numerically disorder chaos in spin glasses, i.e. the
  sensitivity of the ground state to small changes of the random couplings.
  Our study focuses on the Edwards-Anderson model in $d=1,2,3$ and in
  mean-field. We find that in all cases, simple scaling laws, involving the
  size of the system and the strength of the perturbation, are obeyed.  We
  characterize in detail the distribution of overlap between ground states and
  the geometrical properties of flipped spin clusters in both the weak and
  strong chaos regime. The possible relevance of these results to temperature
  chaos is discussed.
\end{abstract}

One of the most spectacular theoretical prediction about the glassy phase of
disordered systems is its generic {\it fragility} to perturbations, in
particular to temperature changes. `Temperature Chaos' (TC), i.e, the chaotic
change of the thermodynamically dominant configurations when temperature is
slightly modified, has been first proposed in the context of the scaling
theory of spin-glasses~\cite{BrayMoore84,FisherHuse86,BrayMoore87,Berker}, and
later extended to other disordered systems, such as pinned elastic
objects~\cite{FisherHuse91}. Although the theoretical situation is well
established in the latter, simpler case~\cite{Marta,RavaJp,Ledou}, the very
existence of
TC in $3d$ spin glasses is still a subject of
controversy~\cite{BilloireMarinari00,KrzakalaMartin02,AspelmeierBray02,MajoFede},
in particular because it has never been directly observed in (static or
dynamic) simulations. It is nevertheless an extremely acute
issue, since TC could be relevant to interpret the spectacular rejuvenation
and memory effects experimentally observed in
spin-glasses\cite{JonasonVincent98,Petra,Sasakius} and in a host of other
materials as well. Arguments for and against the relevance of TC scenario have
been put forward
in~\cite{YoshinoLemaitre00,BDHV,BerthierViasnoff02,BB,GhostStory,BY}.  One
reason for the long standing debate is that if TC occurs at all in
spin-glasses, it does so on very large length scales, much larger than those
accessible to present computers~\cite{KrzakalaMartin02,AspelmeierBray02}.
However, it has been argued that TC can manifest itself even on small length
scales, but only for rare regions of space~\cite{Marta}.  This {\it weak chaos
  regime} has in fact been argued to account in a quantitative way for
experimental results~\cite{GhostStory}. It is therefore particularly important
to clarify the chaotic behavior of spin glasses in that regime. Our strategy
is to study a similar, but much stronger effect~\cite{Hajime04}: the chaotic
dependence of the spin-glass ground state when the {\it couplings} between the
spins are slightly modified.  This is referred to as disorder chaos (DC).
Renormalization group arguments suggest that the two effects are in fact
deeply related and characterized by the same universal scaling
function~\cite{BrayMoore87,FisherHuse91,Marta,RavaJp}; only prefactors differ,
to make the chaos length scale much larger in the TC case. In this letter, we
follow early studies in $2d$~\cite{RiegerChaos,YoungChaos}, and study
numerically DC (at zero temperature) in finite dimensional and mean field
spin-glasses. A precise characterization of {\it both} weak and strong chaos
regime is obtained, even using moderately small system sizes. We discuss in
detail the geometrical aspects of DC, and show that the predictions of the
scaling theory are in agreement with our numerical results, although some of
the assumptions of the droplet theory must probably be revised, a conclusion
already advocated in previous papers~\cite{TNT,Other}. If TC in spin-glasses
is indeed in the same universality class as DC, our results could shed light
on the physics at play when the temperature is slightly changed in
experiments. They also provide a useful guide to interpret numerical
simulations on temperature chaos, which appear to be confined to the weak
chaos regime.

\section{Models and observables}
We work with the Edwards-Anderson ($EA$) Hamiltonian
\be {\cal{H}}=\sum_{i,j} J_{ij}S_i S_j
\label{EA_H}
\ee in finite dimension $d=2$ and $d=3$ with $N=L^d$ Ising spins, and use
periodic boundary conditions. We also study Ising spin glasses on a random
graph (or Bethe Lattice) of fixed connectivity $z=3$. This system is known to
behave like mean field models~\cite{Cavity} and in particular to display
Replica Symmetry Breaking~\cite{MPV}. In order to study disorder chaos, we
consider two copies of the system $\{S^1\},\{S^2\}$ and we modify the couplings $J_{ij}$
in copy $2$ as:
\be 
J_{ij} \rightarrow J_{ij}'=\frac{J_{ij}+x_{ij}\Delta  J}{\sqrt{1+\Delta J^2}},
\label{Bond_Changed}
\ee 
where $x_{ij}$ is a Gaussian random variable with zero mean and unit
variance. With this definition, the original and the modified couplings share
the same distribution. We compute the ground states of both systems
using an optimization method called the Genetic Renormalization
Algorithm~\cite{Algo}. We were able to compute $2000$ instances for different
$\Delta J$s up to $L=60$ in $2d$, $L=10$ in $3d$ and $N=448$ on random graphs.
We will concentrate our analysis on the following observables: 
\be
q=\Big|{\sum_{i=0}^N S^1_i S^2_i}\Big| \qquad\tx{and}\qquad C(r,L) = {\sum_{i=0}^N
S^1_iS^1_{i+r} S^2_iS^2_{i+r}},
\label{def_q} 
\ee 
where $q$ is the absolute value of the {\it spin overlap} between replicas
and $C$ is the {\it four-point, two-replicas correlation function}. The correlation function $C$ therefore 
measures the similarity between the relative orientation of spins a distance $r$ apart
in the original and perturbed system. We will consider quantities averaged
over disorder: $\langle C(r,L)\rangle$ and $\langle q(L)\rangle$, as
well as the distribution of overlaps $P_{L,\Delta J}(q)$.

\section{Scaling predictions}

Following the early arguments of Bray and Moore~\cite{BrayMoore87}, one can
derive several scaling predictions.  Consider the original system with
unperturbed bonds; its ground state configuration(s) are ${\cal{C}}_1$ (and
$-{\cal{C}}_1$). According to the droplet theory, the low lying excitations
are obtained by flipping connected, compact clusters of spins (droplets). A
droplet of size $\ell$ has a fractal surface of dimension $d_s<d$, and its
excitation energy $E> 0$ is distributed as $P(E,\ell) = \ell^{-\theta} \rho
(E\ell^{-\theta})$, where $\rho(x)$ is a scaling function assumed to be non
zero at $x=0$ and which decays to zero for large $x$. The energy exponent $\theta$ is
argued on general grounds to be such that $0 < \theta \leq d_s/2$.
Such a scaling law means that typical droplets of size $\ell$ have an energy
that grows as $\ell^{\theta}$, but that it is still possible to find {\it rare
  excitations} of size $\ell$ with energy $O(1)$. The probability to find such
excitations however decays with $\ell$ as $1/\ell^{\theta}$.  Consider now a
droplet of size $\ell$ in this system, corresponding to a configuration
${\cal{C}}_2$. When the bonds are changed the excess energy of the droplet
also changes. Obviously, this energy comes only from the bonds that differs in
${\cal{C}}_1$ and ${\cal{C}}_2$, and therefore only from the bonds which are
on the {\it surface} of the droplet, the number of which is $\propto
\ell^{d_s}$.  The contribution to the energy of a random perturbation of
$O(\Delta J)$ is thus the sum of $\propto \ell^{d_s}$ independent
random variables with random signs. i.e. a term of order $\pm \Delta J
\ell^{d_s/2}$. If the energy gained by ${\cal{C}}_2$ is larger than its
original excitation energy $\ell^\theta$, then ${\cal{C}}_2$ becomes the new
ground state of the system. If $\theta < d_s/2$, this 
surely happens
for large enough droplets, beyond the 
overlap length 
defined as: \be \ell_c(\Delta J) = \frac{1}{\Delta J^{1/\xi}} \qquad
\tx{with}\qquad \xi=\frac{d_s}{2}-\theta.
\label{len}
\ee
When considering a system-size droplet (i.e. $\ell=L$), this argument suggests that the overlap in Eq.(\ref{def_q}) 
obeys the following scaling law:
\be 
\langle q(\Delta J,L)\rangle = F(L/\ell_c) = F(\Delta J^{1/\xi} L),
\label{scal_Q}
\ee 
where $F(x)$ is a certain scaling function that we now discuss in both the strong and
weak chaos limit.  For small sizes or for small $\Delta J$s, such that $L \ll \ell_c$, 
the perturbation induced by the change in couplings is quite small so that typical droplets do not flip. 
However, there is still a non zero probability to find a large excitation with exceptionally low excess
energy~\cite{Marta,Petra}. The probability to find a rare droplet of size $L$, with an energy less than 
$\Delta J L^{d_s/2}$ is $p(\Delta J,L) \propto L^{-\theta} \int_{0}^{\Delta J L^{d_s/2}} \rho(EL^{-\theta}) dE \sim 
\rho(0) \Delta J L^{\xi}$ (if $\rho(0)$ is finite). If the ground state of the perturbed sample has 
$\propto L^d$ spins flipped with respect to ${\cal C}_1$ with probability $p$, then $1- \langle q 
\rangle \propto \rho(0) \Delta J L^{\xi}$ in the weak chaos limit. In the opposite regime (large size or large $\Delta J$), 
where $L \gg \ell_c$, ${\cal{C}}_1$ and ${\cal{C}}_2$ should become almost independent at scales larger
than $\ell_c$, and the residual overlap should be of order $(\ell_c/L)^{d/2}$. Therefore, one expects:
$F(x) \approx 1 - a x^\xi$ for $x \ll 1$ and $F(x) \approx b/x^{d/2}$ for $x \gg 1$, where $a$ and $b$ are 
constants of order unity. This argument, as we show below, can in fact be generalized to estimate the full
distribution of overlap $P_{L,\Delta J}(q)$, and not only its mean value.

\section{Numerical data}
\begin{figure}
  \includegraphics[width=0.49\textwidth]{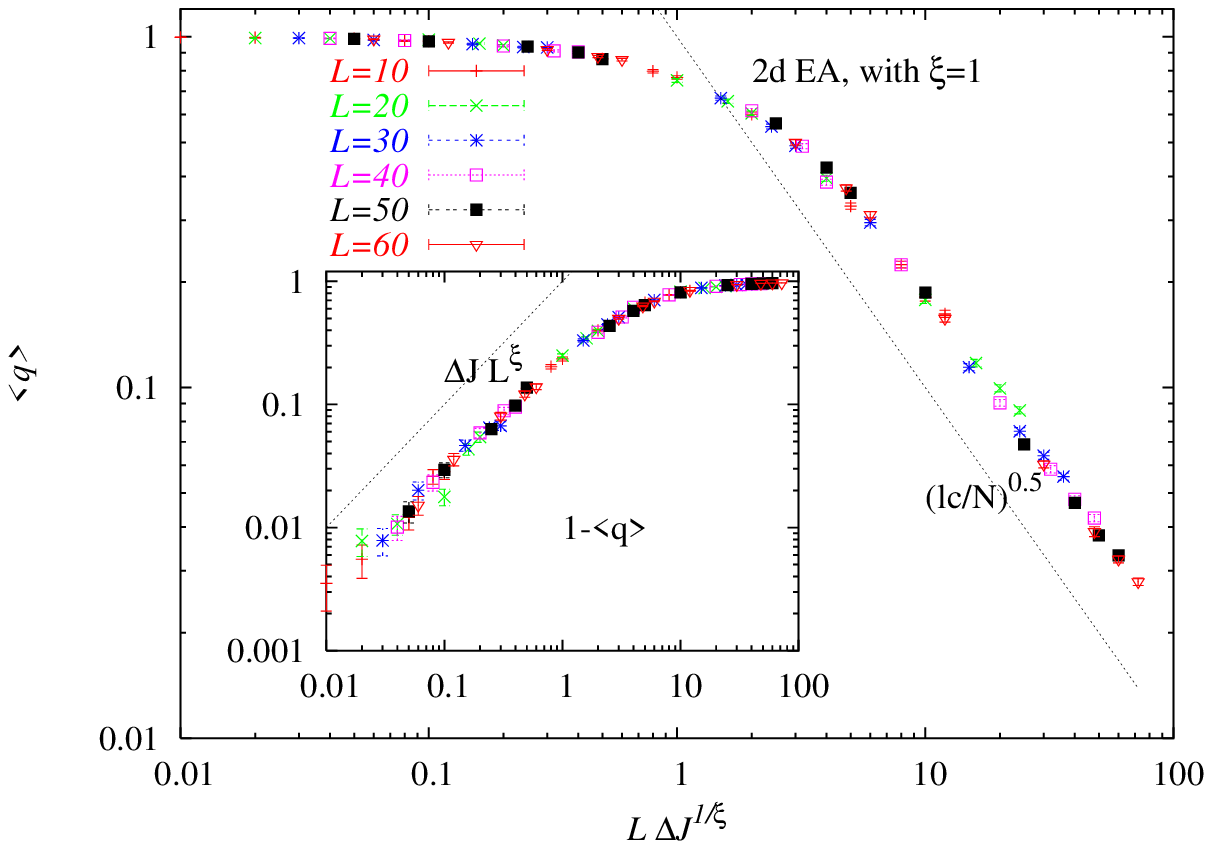}
  \includegraphics[width=0.49\textwidth]{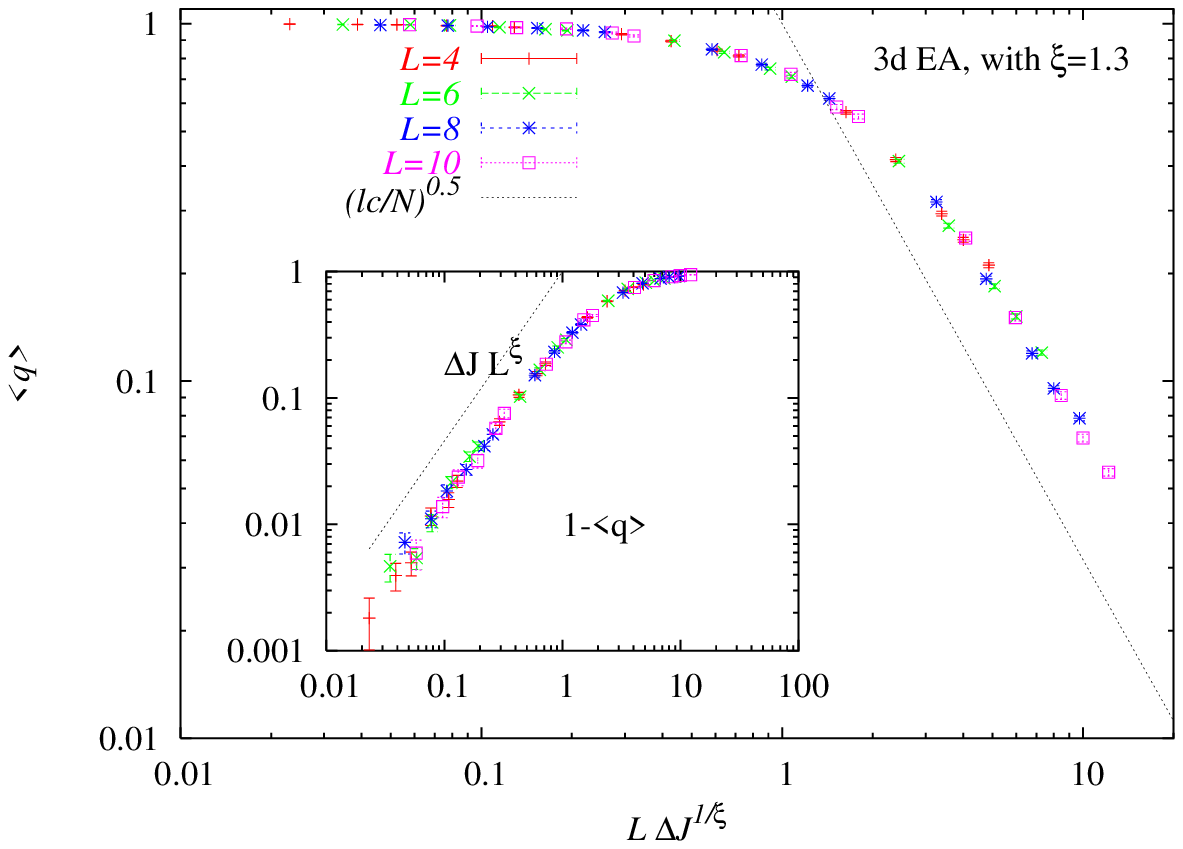}
  \includegraphics[width=0.49\textwidth]{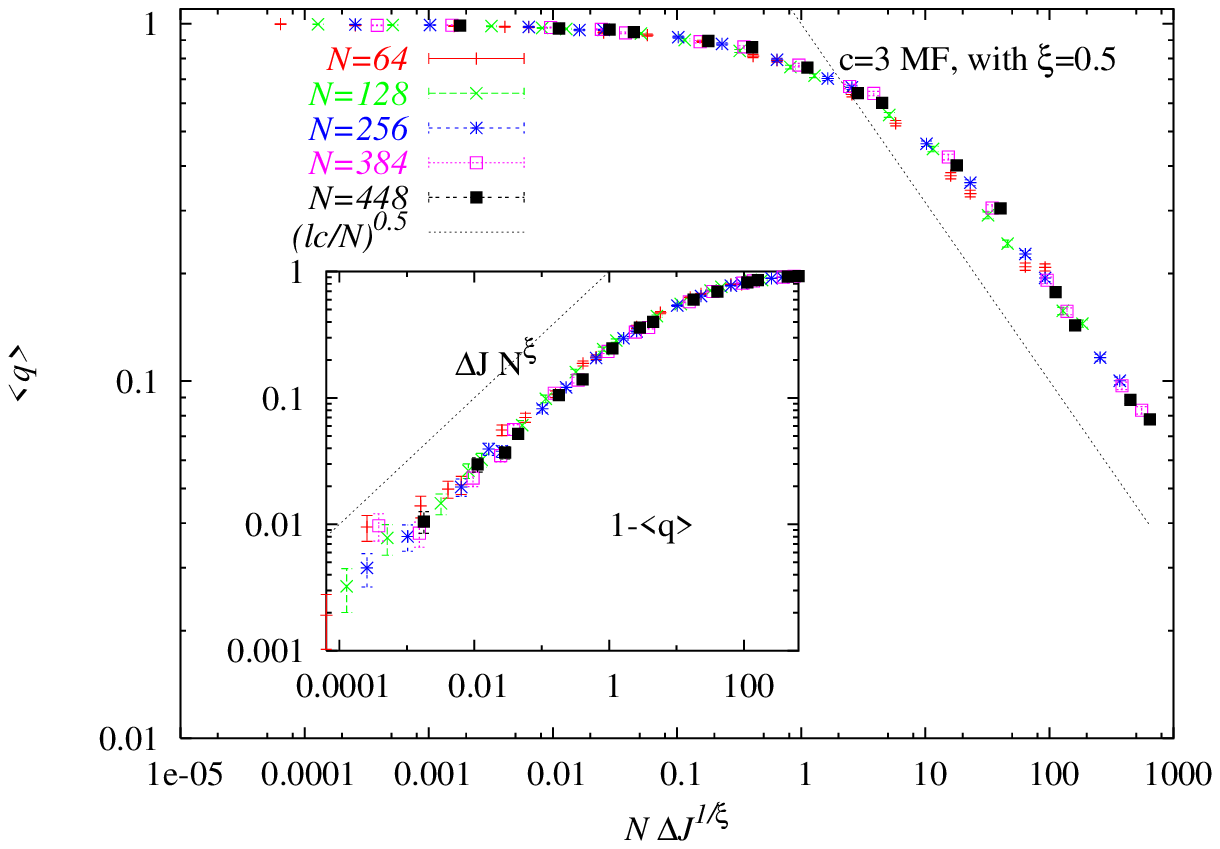}
  \includegraphics[width=0.49\textwidth]{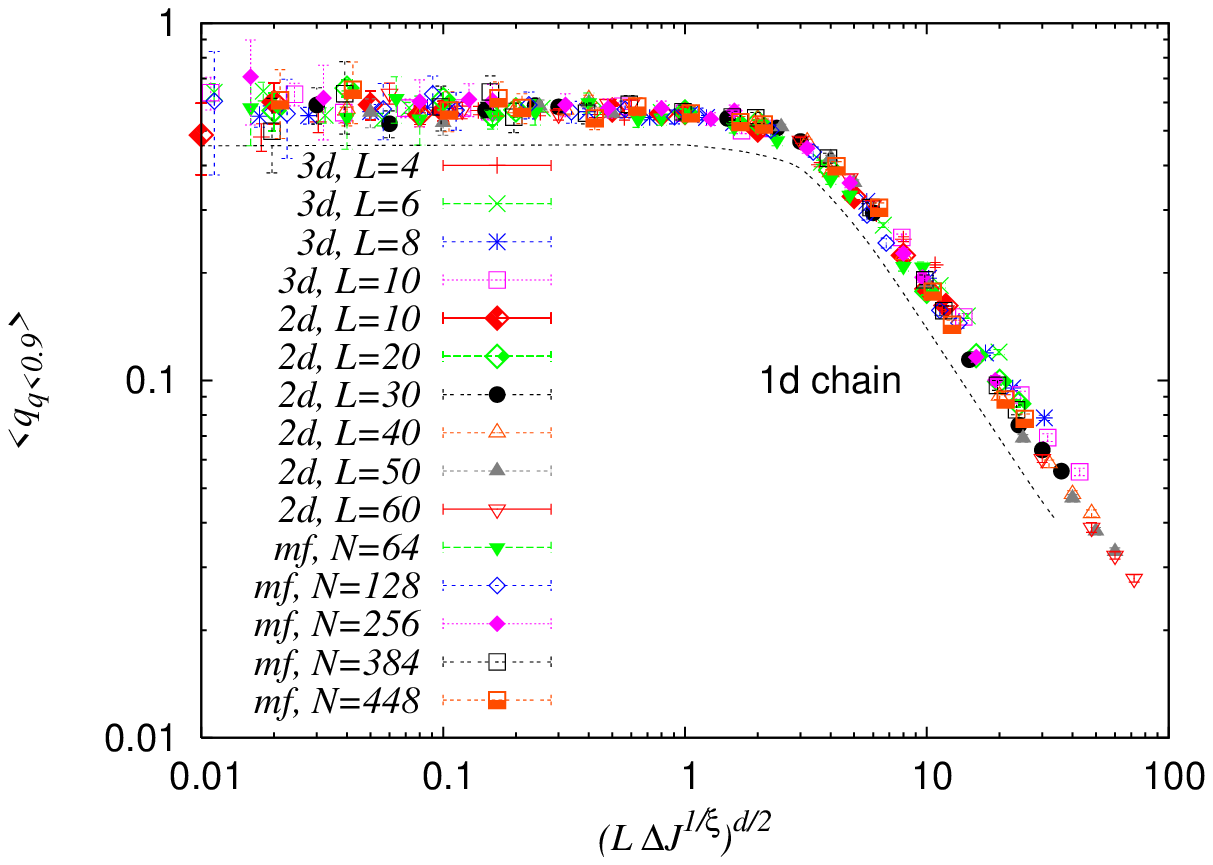}
\caption{Overlap $\langle q(\Delta J,L)\rangle$ versus $L \Delta J^{1/\xi} \equiv L/\ell_c$ in the
  $2d$ and $3d$ $EA$ (top left and right) models and, for a random regular
  graph of connectivity $z=3$, as a function of $N \Delta J^{1/\xi}$ (bottom
  left).  We find, as expected, $\langle q \rangle \propto \sqrt{\ell_c^d/N}$
  in the strong chaos regime and $1-\langle q \rangle \propto \Delta J
  L^{\xi}$ in the weak chaos regime (see insets).  Bottom right: Restricted
  overlap (Eq.(\ref{restricted})) for all models.}
\label{f.1}
\end{figure}
Before turning to a numerical test of these predictions, we first review
existing results on DC. The $1d$ chain can be exactly
solved~\cite{BrayMoore87}, and the $2d$ model was extensively studied by
numerical simulations~\cite{RiegerChaos,YoungChaos}. In both cases,
Eq.(\ref{scal_Q}) is obeyed. There also exist very convincing Monte Carlo
studies in $4d$ for both $T=T_c$ and $T<T_c$~\cite{Hajime04}: these again show
that DC is well described by Eq.(\ref{scal_Q}). There are however no test in
$3d$ nor in mean field (MF), nor systematic investigation of the shape of the
scaling function $F(x)$. Furthermore, nothing is known about the shape of
$P(q)$. We show our results for $\langle q \rangle$ in Fig.  \ref{f.1} for
$2d$, $3d$ and MF. The scaling relations (\ref{scal_Q}) works perfectly in all
three cases, including MF.  The values found for $\xi$ are also in good
agreement with known results for $\theta$ and $d_s$ in the three models.  In
particular: 1) It is widely accepted that the $2d$ model is described by the
droplet theory with $\theta \approx -0.29$ and $d_s \approx 1.3$~\cite{2D},
which agree very well with the value $\xi=1$ that gives the best collapse of
our data in Fig. \ref{f.1}. 2) Mean field systems can also, in some sense, be
described by a droplet theory with $d=d_s$ and $\theta=0$~\cite{FloParisi}, so
that, using $N=L^d$, the scaling variable becomes $\Delta J \sqrt{N}$.  This
is also working very well, demonstrating that, although the structure of mean
field models is very complex~\cite{MPV}, simple scaling arguments are
sometimes enough to understand its behavior~\cite{FloParisi,KrzakalaMartin02}
(note that our definition $\xi$ slightly differs in mean field and finite
dimensional systems). 3) In $3d$, the best collapse is obtained for $\xi
\approx 1.3$. The standard droplet theory however gives $\theta \approx 0.2$
and $d_s \approx 2.6$, which would lead to $\xi_{dr} \approx 1.1$. Mean-field
theory, on the other hand, predicts $\theta =0$ and $d_s=d=3$ so that
$\xi_{mf}=1.5$.  Both theories fail to reproduce the value of $\xi$ favored by
our data (at least for the small sizes we are able to deal with). However, for
these small lattice sizes, the $3d$ EA model is best described by an
intermediate, so-called TNT scenario~\cite{TNT}, where $\theta \approx
0$~\cite{TNT,Other} and $d_s \approx 2.6$~\cite{TNT}.  This is perfectly
consistent with the value $\xi \approx 1.3$ found in Fig. \ref{f.1}.  Fig.
\ref{f.1} also shows, in various insets, that the asymptotic behavior of
$F(x)$ surmised above is correct.

\begin{figure}
  \includegraphics[width=0.48\textwidth]{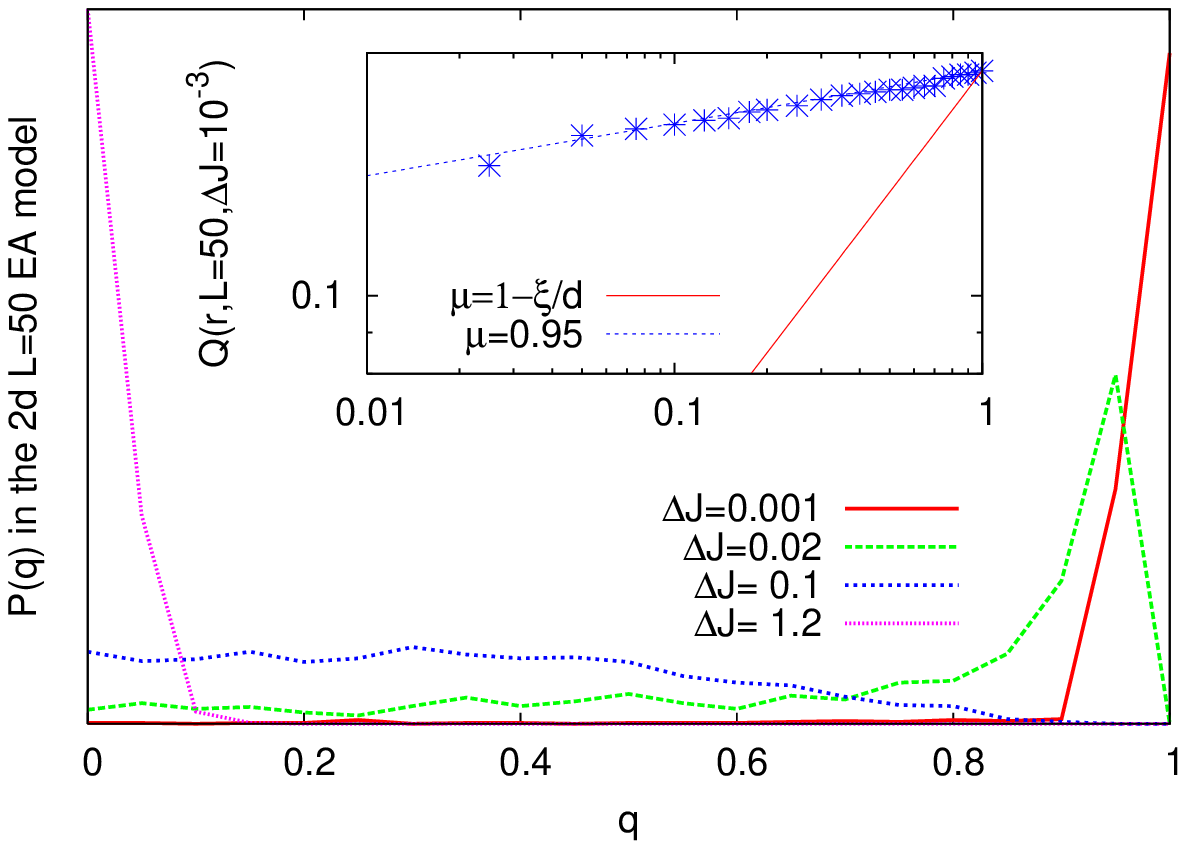}
  \includegraphics[width=0.48\textwidth]{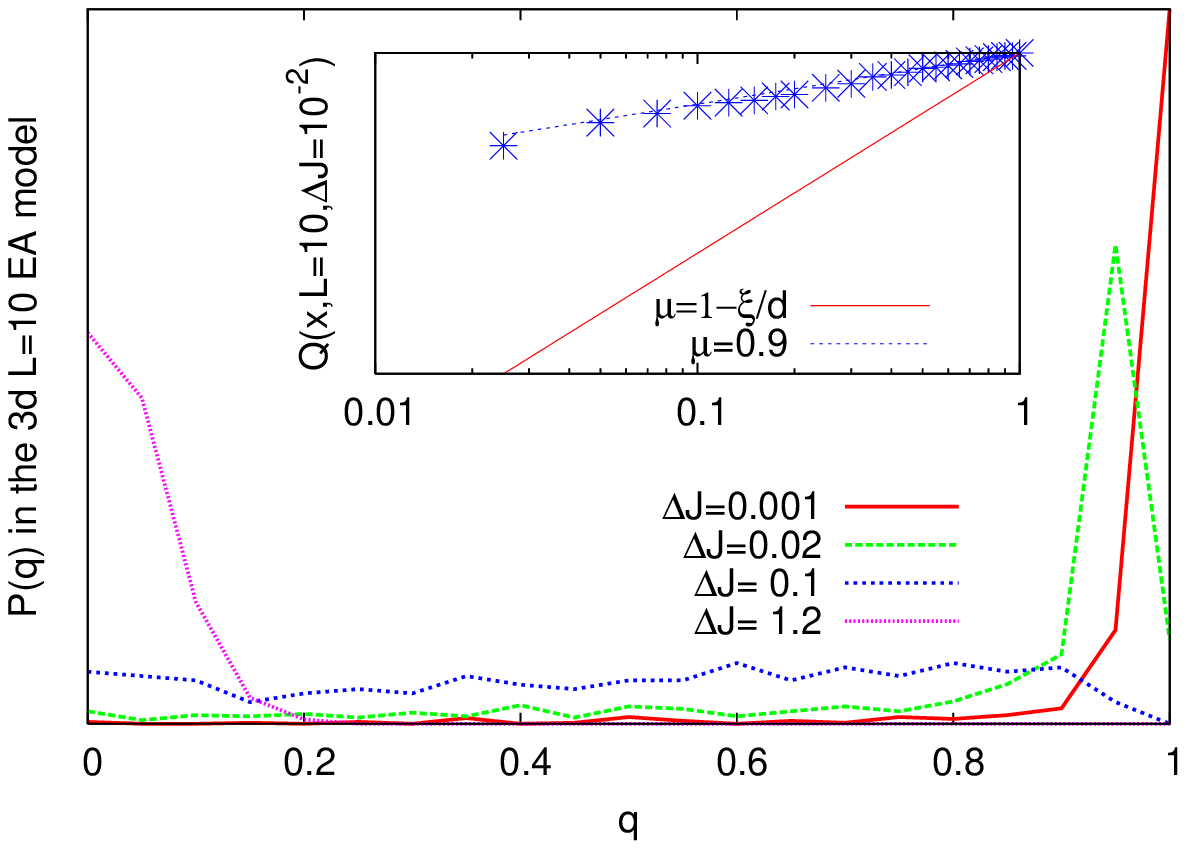}
\caption{The overlap distribution $P(q)$ between ground states in $2d$ (left)
  and $3d$ (right) systems for large size. In inset, $Q(x)=\int_{0}^x
  \widetilde P(r)dr$, where $r=1-q$ for $\Delta J=10^{-3}$(2d) and $\Delta
  J=10^{-2}$(3d).  $Q(x)$ behaves quite differently from the droplet theory
  prediction $Q(x) \propto x^{1-\mu}$, with $\mu={1-\xi/d}$.}
\label{f.3}
\end{figure}

\section{Overlap distribution, weak and strong chaos regime}

We would like to go beyond the analysis of the average overlap to get an
understanding of the distribution of events and their geometry. It is particularly important
to study directly the weak chaos limit since two scenarios are in principle possible:
either a small perturbation typically causes a small rearrangement, i.e., with probability of 
order unity, $O(\Delta J L^\xi)$ spins flip; or a small perturbation typically leads to no 
rearrangement at all, except in rare cases, where if flips $O(L^d)$ spins with probability $O(\Delta J L^\xi)$.
A useful quantity to consider in numerical work is the following restricted average overlap, where we exclude
the samples where almost no change is induced: 
\be
\langle q \rangle _{q<0.9} = \frac{\int_{0}^{0.9} q P(q) dq}{\int_{0}^{0.9} P(q) dq}.
\label{restricted} 
\ee If $P(q)$ is peaked around its average value, $\langle q \rangle _{q<0.9}$
should be close to $0.9$ for small $\Delta J L^\xi$, and progressively
decrease away from that value. If on the other hand, rare events are dominant,
$\langle q \rangle _{q<0.9}$ should be significantly smaller than $0.9$ and
constant as long as $\Delta J L^\xi$ is less than unity. Furthermore, if rare
events induce a complete reshuffling of the spin configuration, as was
conjectured in~\cite{Petra}, then $\langle q \rangle _{q<0.9} \approx 0$.
Fig. \ref{f.1} (bottom right) unambiguously shows that the second scenario
holds: in the weak chaos limit, $\langle q \rangle _{q<0.9}$ is found to be
constant for $\Delta J L^\xi$ small, and close to $0.57$, very different both
from $0.9$ and from $0$. The full distribution $P(q)$ is plotted in Fig.
\ref{f.3} for different values of $\Delta J$. It is clear that, although for
infinitesimal chaos almost all samples lead to $q=1$ and for large chaos to
$q=0$, the distribution in the weak chaos regime is not simply a two-peak
function with weight around $q=0$ and $q=1$. This means that in the rare cases
where the perturbation is relevant, the new ground state typically retains a
significant ``backbone'' of the previous ground state. This is illustrated by
the $1d$ chain case. In the weak chaos limit, only the weakest bond is broken
by the perturbation, so the overlap $q$ is uniformly distributed in $[0,1]$
(in which case $\langle q \rangle_{q<0.9} = 0.45$, see Fig. \ref{f.1}). For $d
> 1$, a droplet like argument predicts $P(q)$ as follows: assume that droplets
of size $\ell$ and $\ell/b$ can be considered as independent for some $b > 1$,
and denote $r=1-q$ the fraction of flipped spins. Defining $\widetilde P(r,L)$
as $P(r,L)$ without the $\delta(r)$ part, one can establish a recursion
equation of the form: $\widetilde P(r,L) = p [f(r)+ b^{d-\xi} \widetilde P(b^d
r,L/b)]$, where $f$ is a regular function.  From this, one deduces that
$\widetilde P(r,L)$ must behave as $r^{-\mu}$, with $\mu=1-\xi/d$.  Using the
values of $\xi$ reported above, we find $\mu=0$ in $d=1$ (which is the exact
result), and $\mu \approx 0.5$ for $d=2,3$ and MF.  This prediction is however
not in agreement with our numerical data (see Fig. \ref{f.3}, insets), which
rather suggests $\mu \approx 1$, i.e. a stronger divergence of $\widetilde
P(r,L)$ at small $r$.  Our data therefore suggests an excess number of small
droplets compared to large ones, possibly related to the findings
of~\cite{Other}, where fractal low energy clusters (corresponding to small
$r$'s) were identified. Along these lines, it is interesting to compare the
system size clusters generated in the weak chaos regime to those obtained
using totally different and more specific methods~\cite{TNT}. Studying the
geometrical properties of the weak chaos droplets, we computed directly the
interface fractal dimension $d_s$ and found, as expected, that $d_s(2d)\approx
1.3$ and $d_s(3d)\approx 2.6$.  We also studied the topological properties of
excitations in $3d$ and found that most of them are spongy, winding around the
lattice. Therefore, the clusters that are flipped in the weak chaos regime are
similar to the large-scale low energy excitations obtained in~\cite{TNT}.

\section{Correlation functions}

The numerical study of the behavior of spatial correlation function between
the two ground states is made difficult because there are now three relevant
lengths in the problem: $L$, $r$ and $\ell_c$ so that $C(r,\Delta J,L)$ may
have non trivial finite size effects (see for instance~\cite{RiegerChaos} for
$2d$ data).  However, we will show that the most important features of the
correlation function can still be fairly well understood. Scaling arguments
suggest that the following form should hold: \be C(r,\Delta J,L)
\label{scal_C}
=\widetilde{C} \((\frac{r}{L},\frac{L}{\ell_c}\)).  
\ee

\begin{figure}
  \includegraphics[width=0.48\textwidth]{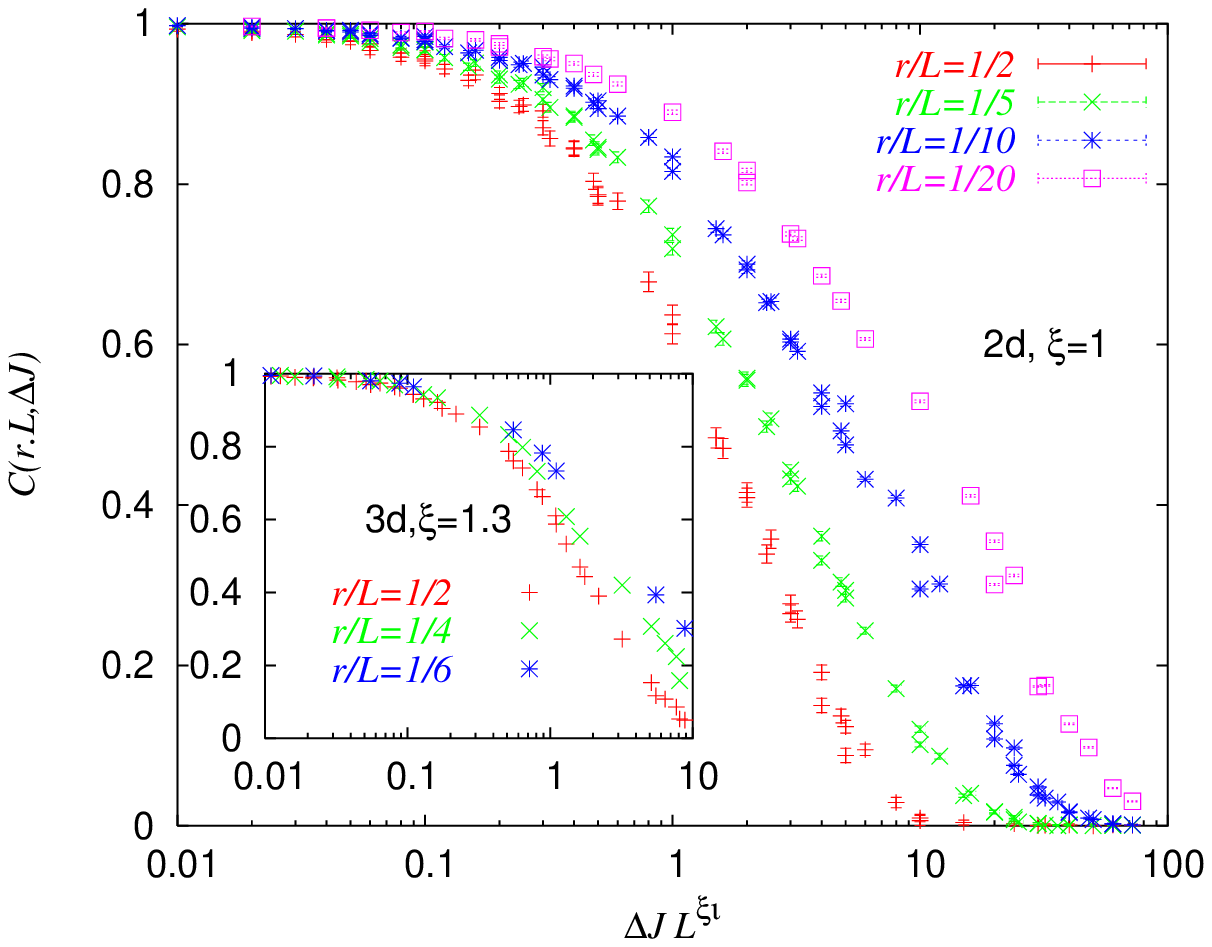}
  \includegraphics[width=0.48\textwidth]{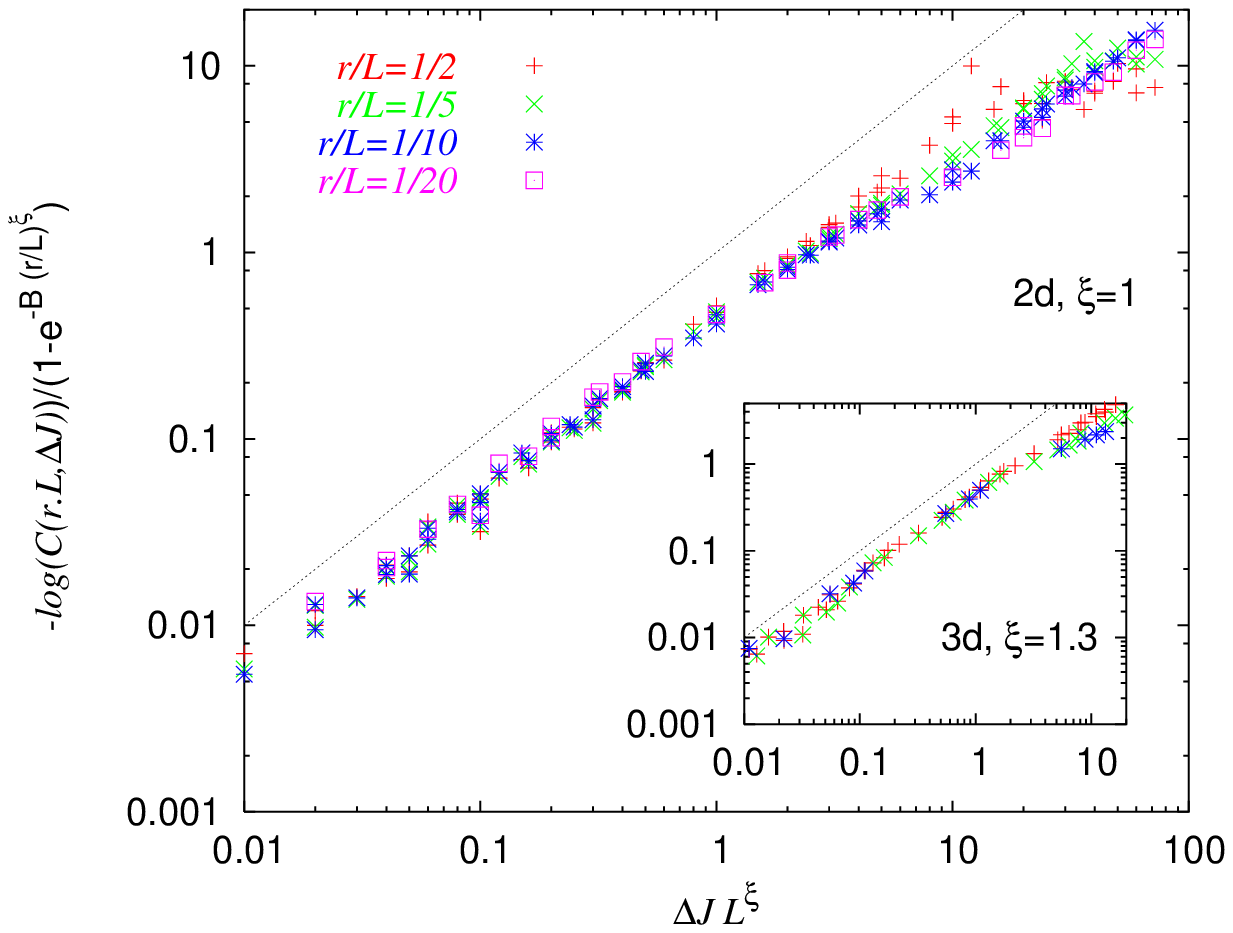}
\caption{Left: Correlation function for size $L=10$ to $L=60$ in the $2d$ EA model
  for different value of the disorder rescaled according to Eq.(\ref{scal_C}).
  Inset: $3d$ data. Right: Collapse of the data using Eq.(\ref{relat_C}); the
  full line is $\((L/l_c\))^{\xi}$ (again $3d$ in the inset).}
\label{f.2}
\end{figure}
Our data in Fig. \ref{f.2} is perfectly compatible with this scaling. Given
this success, one would like to go beyond Eq. (\ref{scal_C}) and propose a specific
form for $\widetilde{C}$. A factorized form does not seem to hold; rather, we found that
the following form leads to a satisfactory collapse of our data both in the weak and
strong chaos regimes (see Fig. \ref{f.2}):
\be
\label{relat_C}
\widetilde{C}(u,v) \approx \exp - \left[A v^{\xi} \left(1-e^{-B
      u^{\xi}}\right)\right], \ee where $A$ and $B$ are fitting constant. In
the limit $L \to \infty$, $u \to 0$ and $\widetilde{C}$ becomes a function of
$w=r/\ell_c$ only, of the form $\exp -AB w^\xi$, which fits well the data, in
particular in the weak chaos regime where it predicts that $1-C(r) \propto
r^{\xi} \Delta J$ as expected from the above arguments applied to droplets of
size $r$. In the long distance regime, this form of $\widetilde{C}$ suggests a
super-exponential decay of the correlation, in agreement with the results on
Migdal-Kadanoff lattices~\cite{SasakiMartin}.

\section{Conclusion}

We have studied numerically disorder chaos in different Ising spin glass
models
finding that, although the $1d$ and $2d$ models have a spin glass
phase only at $T=0$, the $3d$ model has a finite temperature transition and
the mean field model is described by a complex hierarchy of states, DC could
be understood in term of simple droplet-like scaling laws.  In particular, we
find that the weak chaos regime is dominated by rare events where large
droplets are overturned, as conjectured in \cite{Marta,Petra,GhostStory},
with however an anomalous proliferation of small droplets. 
We also find once again strong indications that the droplets are not
compact objects. 
Our study provides precise
predictions that can be tested even on small spin-glass samples, to check
whether temperature chaos is or not in the same universality class as DC. It
would be particularly interesting to reanalyze in this spirit the existing
data of~\cite{BilloireMarinari00} for $3d$ and mean field models.  If the same
scaling functions are obtained, this would constitute compelling evidence
for the presence of a chaotic temperature dependence in spin glasses.  It is
however possible that more complicated scenarii hold, for example temperature
chaos but with continuously varying exponents. As far as experiments are
concerned, our finding of a proliferation of small overturned droplets could
play an important role in the quantitative interpretation of rejuvenation and
memory effects in spin-glasses.

\acknowledgments We thank L. Berthier, T. J\"org, O. Martin, M. Sasaki, R. da
Silveira, E. Vincent, O. White and H. Yoshino for discussions on these issues.


\begin{thebibliography}{0}

\bibitem{BrayMoore84}
A.~J. Bray and M.~A. Moore, J. Phys. C Lett. {\bf 17},  L463  (1984).

\bibitem{FisherHuse86}
D.~S. Fisher and D.~A. Huse, Phys. Rev. Lett. {\bf 56},  1601  (1986).

\bibitem{BrayMoore87}
A.~J. Bray and M.~A. Moore, Phys. Rev. Lett. {\bf 58},  57  (1987).

\bibitem{Berker} S.R. McKay, A.N. Berker, and S. Kirkpatrick, Phys. Rev. Lett.
  {\bf 48}, 767 (1982).

\bibitem{FisherHuse91}
D.~S. Fisher and D.~A. Huse, Phys. Rev. B {\bf 43},  10728 (1991).

\bibitem{Marta} M.~Sales and H.~Yoshino, Phys. Rev. E {\bf 65}, 066131 (2002).

\bibitem{RavaJp} R. A. da Silveira and J-P. Bouchaud, Phys. Rev. Lett. {\bf
    93}, 015901 (2004).

\bibitem{Ledou} P. Le Doussal, 
cond-mat/0505679.

\bibitem{BilloireMarinari00} A. Billoire and E. Marinari, J. Phys. A {\bf 33},
  L265 (2000).

\bibitem{KrzakalaMartin02}
F. Krzakala and O.~C. Martin, Eur. Phys. J. B. {\bf 28},  199  (2002).

\bibitem{AspelmeierBray02} T. Aspelmeier, A.~J. Bray, and M.~A. Moore, Phys.
  Rev. Lett {\bf 89}, 197202 (2002).
  
\bibitem{MajoFede} A. Maiorano {\it et al.} cond-mat/0409577.  M. Picco {\it
  et al}, Phys. Rev. B {\bf 63}, 174412 (2001).

\bibitem{JonasonVincent98}
K. Jonason {\it et~al.}, Phys. Rev. Lett. {\bf 81},  3243  (1998).

\bibitem{Petra} P. E. J\"onsson {\it et al.}  Phys. Rev. Lett. {\bf 89}, 097201
  (2002) and Phys. Rev. Lett. {\bf 90}, 059702 (2003).
  
\bibitem{Sasakius} M. Sasaki, V. Dupuis, J. P. Bouchaud, E. Vincent, Eur. Phys. J. {\bf B29}, 469 (2002).

\bibitem{YoshinoLemaitre00}
H. Yoshino, A. Lemaitre and J.-P. Bouchaud, Eur. Phys. J. B. {\bf 20},  367  (20
00).

\bibitem{BDHV} J. P. Bouchaud, V. Dupuis, J. Hammann, E. Vincent, Phys. Rev. B {\bf 65}, 024439 (2001). 

\bibitem{BerthierViasnoff02} L. Berthier {\it et~al.}, in {\em Les Houches
    2002 session LXXVII}, edited by J.-L. Barrat, J. Dalibard, J. Kurchan, and
  M. Feigel'man (Springer, Berlin, 2003), cont-mat/0211106.
  
\bibitem{BB} L.~Berthier and J.P.~Bouchaud, Phys. Rev. B {\bf 66}, 054404
  (2002).
  
\bibitem{GhostStory} P. E. J\"onsson, R. Mathieu, P. Nordblad {\it et al.}
  Phys. Rev. B {\bf 70}, 174402 (2004).

\bibitem{BY} L.~Berthier and A. P. Young, Phys. Rev. B {\bf 71}, 214429 (2005)
  
\bibitem{Hajime04} M.~Ney-Nifle, Phys. Rev. B 57, 492 (1998);~M. Sasaki {\it
    et al.}, cond-mat/0411138 (2004).


\bibitem{RiegerChaos}
H. Rieger, L. Santen, U. Blasum, M. Diehl, M. J\"unger, J. Phys. A {\bf 29}, 3939 (1996).

\bibitem{YoungChaos}
M.~Ney-Nifle and A.P.~Young,  J. Phys. A, {\bf 30}, 5311 (1997).

\bibitem{TNT} 
F. Krzakala and O.C.~Martin, Phys. Rev. Lett {\bf 85}, 3013  (2000).  
M. Palassini and A.P.~Young, Phys. Rev. Lett. {\bf 85}, 3017 (2000).
J. Houdayer {\it et al.}, Eur. Phys. J. B {\bf18}, 467 (2000).

\bibitem{Other} J. Lamarcq, J. P. Bouchaud, O. C. Martin and M. M\'ezard,
  Europhys. Lett. {\bf 58}, 321 (2002).

\bibitem{Cavity}
M. M\'ezard and G. Parisi, Eur. Phys. J. B {\bf 20}, 217 (2001)

\bibitem{MPV} M. M{\'e}zard, G. Parisi, and M. A. Virasoro, {\it Spin-Glass
    Theory and Beyond}, Lecture Notes in Physics Vol. 9 (World Scientific,
  Singapore,1987).
  
\bibitem{Algo} J. Houdayer and O. C. Martin, Phys. Rev. Lett. {\bf 83}, 1030
  (1999).



\bibitem{2D} M. Palassini and A. P. Young, Phys. Rev. B {\bf 60}, R9919
  (1999); A. Middleton, Phys. Rev. Lett. {\bf 83}, 1672-1675 (1999), A. K.
  Hartmann {\it el al.}, Phys. Rev. B {\bf 66}, 224401 (2002)


\bibitem{FloParisi} F. Krzakala and G. Parisi, Europhys. Lett. {\bf 66},
  729-735 (2004).
  
\bibitem{SasakiMartin} 
M. Sasaki, O.~C. Martin, Phys. Rev. Lett. {\bf 91}, 097201 (2003).

\end{thebibliography}
\end{document}